\documentclass[twocolumn,showpacs,amsmath,amssymb]{revtex4}

\newcommand{\be}{\begin{equation}}  
\newcommand{\ee}{\end{equation}}

\newcommand{\bea}{\begin{eqnarray}}           
\newcommand{\eea}{\end{eqnarray}} 
\newcommand{\ba}{\begin{align}}
\newcommand{\ea}{\end{align}}
\def\de{\partial}
\def\be{\begin{equation}}
\def\ee{\end{equation}}
\def\beq{\begin{eqnarray}}
\def\eeq{\end{eqnarray}}
\newcommand{\beqn}{\begin{eqnarray*}}
\newcommand{\eeqn}{\end{eqnarray*}}
\def\lm{{\ell m}}

\def\l{{\ell }}

\def\bI{\hbox{$\,I\!\!\!\!-$}}

\def\IL{\relax{\rm I\kern-.18em L}}

\usepackage{colordvi}
\usepackage{latexsym}
\usepackage{amssymb}
\usepackage{amsfonts}
\usepackage{amsmath}
\usepackage[dvips]{graphicx}

\begin{document}

\title{Gravitational waves from oscillating accretion tori: \\
Comparison between different approaches}

\author{Alessandro Nagar$^{1,2}$ 
       , Jos\'e A. Font$^{1}$
       , Olindo Zanotti$^{1}$
      and Roberto De Pietri $^{3}$}

\affiliation{ $^{1}$Departament d'Astronomia i Astrof\'{\i}sica,
                     Universitat de Val\`encia, Edifici d'Investigaci\'o,
                     Dr.~Moliner 50, 46100 Burjassot (Val\`encia), Spain\\      
                    $^{2}$ Dipartimento di Fisica, Politecnico di Torino,
                     Corso Duca degli Abruzzi 23, 10129
                     Torino, Italy \\                          
                $^{3}$ Dipartimento di Fisica, Universit\`a di Parma, 
                 Parco Area delle Scienze 7a, 43100, Parma, Italy}

\date{\today}

\begin{abstract}
 Quasi-periodic oscillations of high density thick accretion disks orbiting 
 a Schwarzschild black hole have been recently addressed as interesting sources 
 of gravitational waves. The aim of this paper is to compare 
 the gravitational waveforms emitted from these sources when computed using 
 (variations of) the standard quadrupole formula and gauge-invariant metric perturbation 
 theory. To this goal we evolve representative disk models using an existing general 
 relativistic hydrodynamics code which has been previously employed in investigations 
 of such astrophysical systems. Two are the main results of this work: First, for 
 stable and marginally stable disks, no excitation of the black hole quasi-normal 
 modes is found. Secondly, we provide a simple, relativistic modification of 
 the Newtonian quadrupole formula which, in certain regimes, yields excellent 
 agreement with the perturbative approach. This holds true as long as back-scattering 
 of GWs is negligible. Otherwise, any functional form of the quadrupole formula yields
 systematic errors $\sim 10\%$.
\end{abstract}


\maketitle
Quasi-periodic oscillations of axisymmetric thick accretion disks 
or tori orbiting around a black hole have been recently addressed as promising 
sources of gravitational waves (GWs)~\cite{zanotti03}. In Ref.~\cite{zanotti03} 
the GW signals were computed by means of the quadrupole stress formula 
within the Newtonian approximation, as it is commonly done in the 
literature~\cite{blanchet90,zwerger97,finn90,dimmelmeier,shibata03}. 
The purpose of the present article is to compare the gravitational 
waveforms computed using the quadrupole formalism with those obtained 
from the theory of (gauge-invariant) gravitational metric perturbations of 
Schwarzschild spacetime, which amounts to solve (in the time domain) the 
inhomogeneous even- (Zerilli-Moncrief) and odd-parity (Regge-Wheeler) 
master equations~\cite{regge57,zerilli70,moncrief74}. Since our interest 
is on the comparison between the two approaches, we shall concentrate on
even-parity perturbations only. We note that similar investigations have 
been reported by~\cite{shibata03} in the context of oscillations of neutron 
stars, and by~\cite{tanaka93} for orbiting particles around a Schwarzschild 
black hole. 

In order to perform such a comparison we use the same physical setup 
of~\cite{zanotti03}, namely a relativistic thick accretion disk
undergoing oscillations induced by suitable initial perturbations of 
an equilibrium configuration. A number of restrictive hypothesis are 
assumed in our computational framework which, however, are valid as 
long as the mass of the disk $\mu$ is much smaller than the mass of 
the black hole $M$. In such a case, the self-gravity of the disk and 
the radiation reaction effects can be neglected. We assume thus that 
the spacetime metric $\bar{g}_{\mu\nu}$ is described by the Schwarzschild 
metric $g_{\mu\nu}$ plus a non-spherical perturbation $h_{\mu\nu}$. 
As a result of the spherical symmetry, $h_{\mu\nu}$ can be 
expanded in seven even-parity and three odd-parity multipoles
\begin{equation}
\bar{g}_{\mu\nu}=g_{\mu\nu}
+\sum_{\l=2}^{\infty}\sum_{m=-\l}^{\l}
\left(h_{\mu\nu}^{\lm}\right)^{(\mathrm{e})}+\left(h_{\mu\nu}^{\lm}\right)^{(\mathrm{o})}\;,
\end{equation}
%
%----------------------------------- TABLE I: the models ----------------------------------------------
\begin{table*}[t]
\caption{\label{label:table1}
Marginally stable (${\rm A, C}$) and stable (${\rm B, D}$) constant angular
momentum thick disks orbiting 
around a Schwarzschild black hole of mass $M=2.5 M_{\odot}$. From left to 
right the columns report the name of the model, the number of radial and 
polar grid zones used in the simulations, the disk-to-hole mass ratio, the 
polytropic constant $\kappa$ of the isoentropic equation of state (EOS) 
$p=\kappa\rho^\gamma$ with $\gamma=4/3$, the value of the specific angular 
momentum $l$, the position of the cusp $r_{\rm cusp}$ and of the center 
$r_{\rm center}$ of the disk, the density at the center $\rho_{\rm c}$, 
the location of the inner ($r_{\rm in}$) and outer ($r_{\rm out}$) disk 
boundaries, the linear size in the equatorial plane $L$, the orbital 
period of the center $t_{\rm orb}$, the value of the potential barrier 
$\Delta W$, and the magnitude of the perturbation parameter $\eta$. The inner 
and outer radii of the computational domain are $r_{\mathrm{min}}=2.1$ (for
all models) and $r_{\mathrm{max}}=18$ (A), 20 (B and C), and 40 (D). 
All radii are given in units of $M$.}
\begin{ruledtabular}
\medskip
\begin{tabular}{ccccccccccccccc}
Model  & $N_r$ & $N_\theta$ &$\mu/M$ & $\kappa$ (cgs)& $l$ & $r_{\rm cusp}$ &$r_{\rm center}$ & $\rho_{\rm c}$ (cgs) & 
$r_{\rm in}$& $r_{\rm out}$ & $\,L$ (km) &  $t_{\rm orb}$ (ms) & $\Delta W$ & $\eta$ \\
\hline
${\rm A}$ &320  & 150 &$0.10$   & $3.01\times 10^{13}$ & 3.75 & 4.835 & 7.720 & $3.58\times10^{13}$ 
& 4.83 & 11.57 &24.87 & 1.66 &  $0$  & 0.1 \\
${\rm B}$ &300  & 150  &$0.02$      & $3.58\times 10^{13}$& 3.75& 4.835 & 7.720 & $1.39\times10^{13}$ 
& 5.81 & 10.32 & 16.65 & 1.66 & $-0.002 $ & 0.01 \\
${\rm C}$ &300  & 150 & $0.10$  & $9.60\times 10^{13}$ & 3.80& 4.576 &8.352  & $1.15\times10^{13}$ 
& 4.57 & 15.89 & 41.76& 1.86 & 0 & 0.01 \\
${\rm D}$ &300  & 150 & $0.01$ & $6.19\times 10^{14}$  & 3.95& 4.107 &9.971 & $2.83\times10^{11}$ 
& 5.49 & 29.08 &87.09 & 2.43 &  $-0.015$ & 0.02 \\
\end{tabular}
\end{ruledtabular}
\end{table*}
%-----------------------------------------------------------------------------------------------------------------------
with the even multipoles transforming as $(-1)^{\l}$ and the odd as $(-1)^{\l+1}$ 
under a parity transformation $(\theta,\varphi)\rightarrow(\pi-\theta,\pi+\varphi)$. 
The presence of a fluid matter disk around the black hole is accounted
through a ``source'' term in the linearized Einstein's equations, 
represented by a stress energy tensor $t_{\mu\nu}$ which can be expanded in 
multipoles as well (see~\cite{nagar04a,nagar04b} for details). By taking suitable 
combinations of the metric multipoles, it is possible to define an 
even-parity $Z^{\lm}$ (Zerilli-Moncrief) and an odd-parity ${\it \Psi}^{\lm}$ (Regge-Wheeler)
gauge-invariant master function. The linearized Einstein's equations for 
the even-parity part reduce to the Zerilli-Moncrief equation in $Z^{\lm}$
\begin{align}
Z_{,tt}^{\lm}-Z_{,r_*r_*}^{\lm}+V_{\l}^{({\rm e})} Z^{\lm}&=S_z^{\lm}\;,
\label{inzerilli}
\end{align}
where $r_*=r+2M\log[r/(2M)-1]$ is the Regge-Wheeler tortoise coordinate and 
\begin{align}
&V_{\l}^{({\rm e})}=\left(1-\dfrac{2M}{r}\right)\\
&\times\dfrac{\lambda(\lambda-2)^2r^3
+6(\lambda-2)^2Mr^2+36(\lambda-2)M^2r+72M^3}
{r^3\left[(\lambda-2)r+6M\right]^2}\nonumber
\end{align}
is the Zerilli potential, with $\lambda=\l(\l+1)$. Explicit expression for 
the source $S_z^{\lm}$, in the gauge-invariant formalism of 
Gerlach-Sengupta~\cite{gerlach79}, can be found in Refs.~\cite{nagar04a,nagar04b}. 
In the wave-zone (i.e. for large $r$) it can be shown that the even-parity GW 
amplitude reads (see e.g.~\cite{nagar04b} and references therein)
\begin{align}
\label{eq:h}
h_+-{\rm i}h_{\times}&=\dfrac{1}{2r}\sum_{\l=2}^{\infty}\sum_{m=-\l}^{\l}
Z^{\lm}\sqrt{\frac{(\l+2)!}{(\l-2)!}}\;_{-2}Y^{\lm}\left(\theta,\varphi\right),
\end{align}
where $_{-2}Y^{\lm}\left(\theta,\varphi\right)$ are the $s=-2$ spin-weighted 
spherical harmonics. In axisymmetry ($m=0$)
\begin{align}
\label{eq:even_signal}
h_+      &= \dfrac{1}{2r}\sum_{\l=2}^{\infty}Z^{\l0}\;W^{\l0}\;,
\end{align}
where $W^{\l0}=Y^{\l0}_{,\theta\theta}-\cot\theta Y^{\l0}_{,\theta}$, and $h_{\times}=0$.
The total energy radiated in GWs is defined by
\begin{align}
\label{energy:Z}
E^{\l0}&=\int_0^{\infty}\!\!\left(\frac{dE}{d\omega}\right)^{\l0}d\omega\nonumber\\
&=\dfrac{1}{64\pi^2}\frac{(\l+2)!}{(\l-2)!}\int_0^\infty \omega^2|\tilde{Z}(\omega,r)|^2d\omega\;,
\end{align}
where we have introduced the Fourier transform $\tilde{Z}(\omega,r)=
\int_{-\infty}^{\infty}e^{-i\omega t}Z(t,r)dt$, with $\omega=2\pi f$ and $Z\equiv 
Z^{\l0}$. 
In the following we focus our discussion on the 
$\l=2$ multipole where $W^{20}=\sqrt{45/(4\pi)}\sin^2\theta$. 
%----------------------------------- FIGURE 1 ---------------------------------------------
\begin{figure*}[t]
\begin{center}
\includegraphics[width=75 mm, height=65 mm]{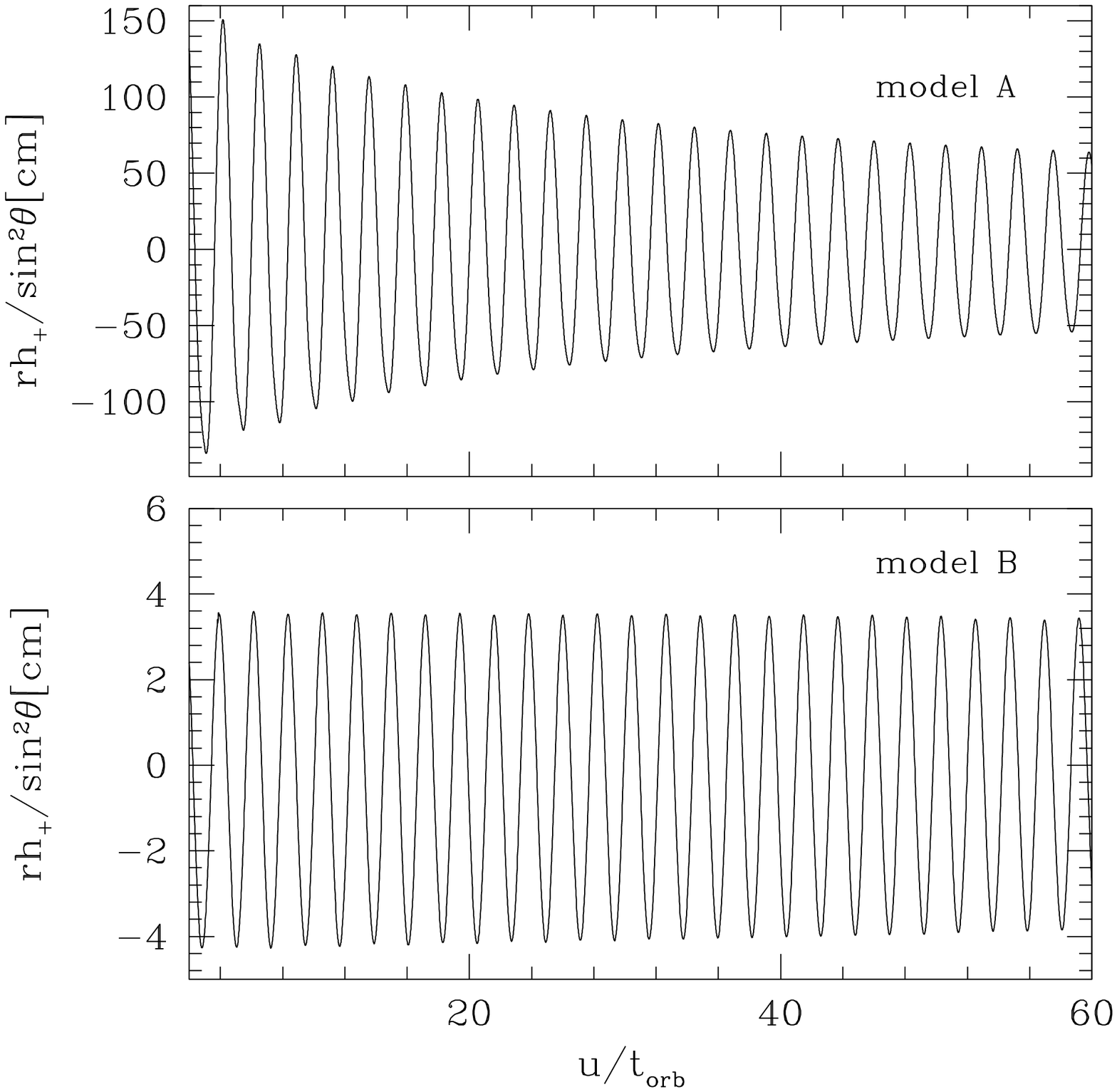}\qquad
\includegraphics[width=75 mm, height=65 mm]{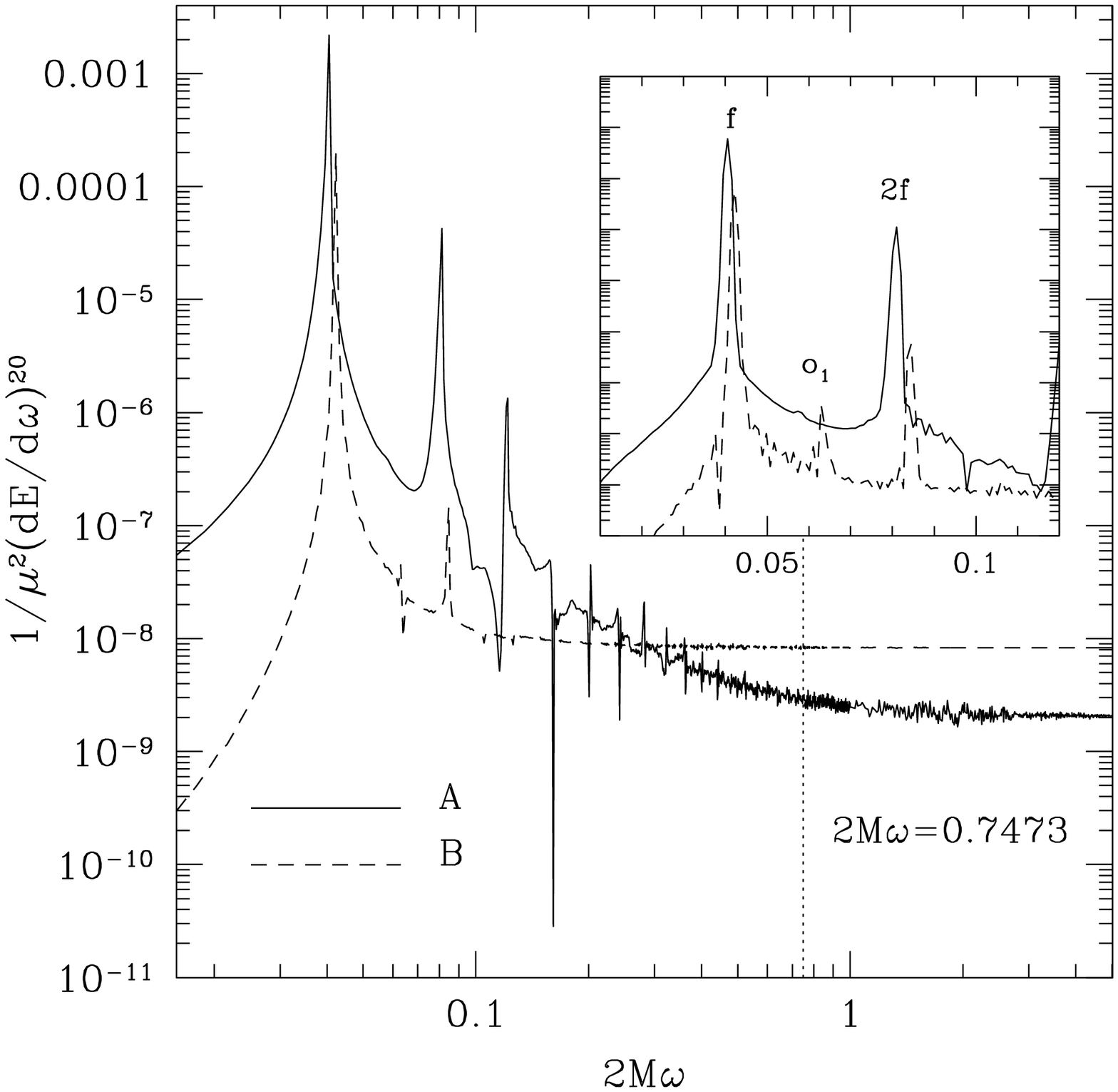}
\caption{
{\it Left panel}: Time evolution of the $\l=2, m=0$ GWs for 
models A and B, computed using the Zerilli-Moncrief equation. The first 
60 periods of a $t=100 t_{{\rm orb}}$ simulation are shown. The oscillatory 
pattern in the waveforms is the response of the quasi-periodic oscillations of the disks. {\it 
Right panel}: Energy spectra for model A (solid line) and B (dashed line). Only 
fluid $p$-modes are visible in the spectra (enlarged in the inset, where a 
Hamming filter  was used) which show no evidence 
of the excitation of the fundamental QNM of the black hole (its frequency is
indicated by a vertical dotted line at $2M\omega = 0.7473$).} 
\label{label:fig1}
\end{center}
\end{figure*}
%-------------------------------------------------------------------------------------------

The equilibrium accretion disks that we use as initial models for our simulations,
and whose hydrodynamics is accounted for in the source terms of Eq.~(\ref{inzerilli}), 
are built by solving the relativistic Bernoulli equation as described in~\cite{font02a}. 
We only recall here that the tori are isoentropic, non-Keplerian (thick) disks with 
a cusp on the equatorial plane, through which matter can accrete onto the black hole 
provided there is some departure from hydrostatic equilibrium. For simplicity, we only 
consider initial models with a uniform radial distribution of the specific angular 
momentum, a fact which guarantees that the radial location of the cusp is always 
smaller than the marginally stable orbit. Despite constant specific angular momentum 
disks do not represent physical tori, being dynamically unstable against nonaxisymmetric 
instabilities, such a simple model is still appropriate for our purpose of testing the 
quadrupole formula (see~\cite{font02a} for details on improved models of equilibrium 
tori). The geometrical structure of the equipotential surfaces of the tori resembles 
that of the Roche lobes of a binary system. In particular, by varying the value of 
the potential barrier $\Delta W=W_\mathrm{in}- W_\mathrm{cusp}$ at the inner edge 
of the disk, it is possible to build either a stable configuration corresponding 
to a torus inside its Roche lobe ($\Delta W<0$), or a torus overflowing its Roche 
lobe ($\Delta W>0$) and therefore accreting onto the black hole. The case $\Delta W=0$ 
corresponds to the marginally stable case. 

In the present work we deal with models which are stable or marginally stable with
respect to the runaway instability (see e.g.~\cite{font02a} and references therein). 
Table~\ref{label:table1} lists the main properties of the models considered. 
The mass of the black hole is $M=2.5 M_{\odot}$, and disks with different disk-to-hole
mass ratio $\mu/M$ are considered. In all cases, but particularly for models B and D, 
the disk represents a small perturbation to the gravitational field of the black hole. 
Models A and B share the same location of the center and the cusp (i.e.~have the same 
equipotential surfaces), but differ in the total mass $\mu$ and in the value of the 
potential gap $\Delta W$. In order to induce periodic oscillations of the disks we 
follow the same procedure of~\cite{zanotti03} and introduce a small radial velocity 
perturbation, parameterizing its strength in the form $v_r = \eta v_r^{\rm sph}$, 
where $v_r^{\rm sph}$ is the (stationary) radial velocity of the spherically 
symmetric low density atmosphere surrounding the torus. The values of the 
perturbation parameter chosen are typically in the interval $\eta\in[0.01,0.1]$. 
This produces low mass accretion rates (see below) and validates the assumption 
of a ``fixed'' background spacetime. The initial data are evolved in time using 
an axisymmetric general relativistic hydrodynamics code based on the conservative 
formulation of the equations, whose accuracy has been assessed in earlier 
investigations~\cite{font02a,zanotti03,nagar04a}. The computational 
$(r,\theta)$-grid used in the simulations has non-uniform radial spacing and 
equidistant polar spacing. Table~\ref{label:table1} reports the number of grid zones 
employed for each model.
 
It needs to be mentioned that setting up the initial data for even-parity perturbations 
requires some care~\cite{nagar04a}, as the initial profile of $Z$ and $Z_{,t}$ should be 
obtained by solving the (linearized) Hamiltonian and momentum constraints with matter 
sources, respectively. When this is not done, an initial unphysical burst of gravitational 
radiation is produced, which may cause the excitation of the black hole quasi-normal 
modes (QNM) and hide physical effects if these were to happen on a comparable timescale.  
However, this is not the case in our simulations as the timescale of the oscillations 
of the disk is much larger than the decay time of the black hole QNM. Hence, once the 
initial data have been evolved for a sufficient time (roughly a couple of orbital periods) 
the gravitational waveforms remain entirely unaffected by the initial GW content.  
In practice, we choose for the initial profile of $Z$ a $\l=2$ time-symmetric stationary 
multipole by solving Eq.~(\ref{inzerilli}) as an ODE with $Z_{,tt}=0$.

The left panel of Fig.~\ref{label:fig1} shows the $\l=2$ gravitational waveforms 
extracted for models A and B through the Zerilli-Moncrief equation (at $r_{{\rm obs}}=400 M$) 
as a function of the observer retarded time $u=t-r_*^{\rm obs}$, in units of the 
orbital period of the center of the disk. Due to the finite location of the observer
a vertical offset is present in the waveforms, induced by the matter sources in 
Eq.~(\ref{inzerilli}). This offset has been corrected for in the figure. The total mass 
accreted onto the black hole at the end of the simulations ($t=100t_{\rm orb}$) is 
$\sim 0.03M$ for model A and $\sim 2.20\times 10^{-4} M$ for model B, which justifies 
the fixed spacetime approximation we use. The waveforms shown in the left panel of 
Fig.~\ref{label:fig1} indicate that the GW signals follow the torus oscillations. 
The decrease of the GW amplitude is determined by the amount of accreted mass. For 
model B the GW amplitude remains almost constant throughout the evolution, while 
for model A, where more mass transfer is possible at the cusp, it decreases by 
$\sim 60\%$ after 100 orbital periods. As becomes evident on the right panel of the 
figure, which shows the corresponding energy spectra for both models, no black hole 
QNM is excited in the process. The spectra have been obtained 
after cutting out the first four periods of oscillation in order to avoid any 
contamination from the early time unphysical burst. At low frequencies (see the 
inset in the right panel of Fig.~\ref{label:fig1}), distinctive peaks are visible, 
which correspond to the fluid $p$-modes of the disk, namely a fundamental mode $f$ 
and a series of overtones~\cite{zanotti03,rezzolla03}. The fundamental $\l=2$ QNM of the black 
hole should lie at $2M\omega=0.7473$ (indicated by a vertical dotted line in the plot), 
but the computed spectra show no hints of this mode. This is consistent with the
results reported in~\cite{nagar04a} for the case of radial infall of quadrupolar
shells, where it was shown that the fundamental QNM of the black hole is only significantly 
excited if the bulk of the maximum density of the (compact) shell reaches the peak 
of the Zerilli potential and if its typical linear dimensions are comparable with 
the width of the potential itself.

%----------------------------------------------
% - Considerations about the quadrupole formula
%----------------------------------------------
The absence of black hole QNM excitation in our simulations suggests that
the gravitational waveforms should also be accurately computed by means 
of the Landau-Lifshitz quadrupole formula. We use our setup to assess the 
validity of the quadrupole formalism through a direct comparison with the 
gauge-invariant GWs extraction shown in Fig.~\ref{label:fig1}. Since the quadrupole 
formula has been derived in a Newtonian framework, a number of approximations 
must be first considered in the transition to a general relativistic context. 
Let us consider a Minkowski background spacetime $\eta_{\hat{\mu}\hat{\nu}}$ 
(${\hat{\mu}}=0,1,2,3$) and an axisymmetric GW source whose physical dimensions 
are much smaller than the reduced gravitational wavelength (i.e. slow-motion approximation).
In such case the only non-vanishing polarization amplitude of 
the wave is $h_+$, which is given by~\cite{MTW,finn90,shibata03}
\begin{align}
\label{eq:h_plus}
h_+ = \frac{3\sin^2\theta}{2r}\,\frac{d^2\bI_{\hat 3\hat 3}}{dt^2}\;,
\end{align}
where
\begin{align}
\label{eq:I_slash}
\bI_{\hat{i}\,\hat{j}}\equiv\int t^{\hat 0\hat 0}\left(x^{\hat{i}}x^{\hat{j}}-
\frac{1}{3}\delta^{\hat i\hat j}r^2\right)d^3x
\end{align}
becomes the reduced matter quadrupole moment tensor in the Newtonian
limit $t^{\hat 0\hat 0}= \rho$ \cite{MTW}, $\rho$ being the rest-mass density of the source. 
Eq.~(\ref{eq:h_plus}) is known as the Standard Quadrupole Formula (SQF)
%---------------------------------------------------------------------------
\begin{figure}[t]
\begin{center}
\includegraphics[width=75mm , height=65 mm]{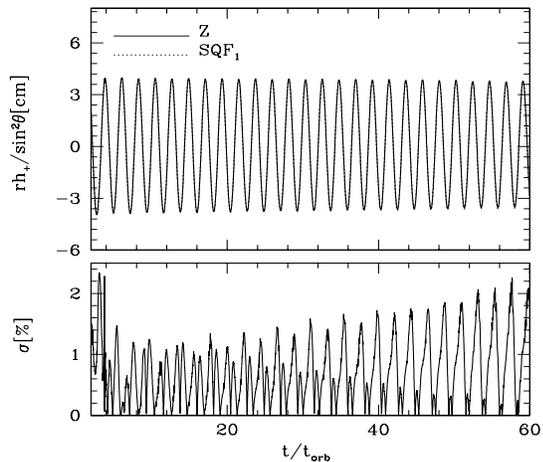}
\caption{
Comparison between the $\l=2$ gravitational waveforms for model B extracted using the
quadrupole formula SQF$_1$ (dotted line) and the gauge-invariant approach (solid
line). The bottom panel shows the relative difference between the two normalized to 
the mean amplitude of the gauge-invariant method.}
\label{label:fig2}
\end{center}
\end{figure}
%---------------------------------------------------------------------------
and can be used to extract GWs in a slow-motion nearly Newtonian regime. 
An extension of the simple Eqs.~(\ref{eq:h_plus}) 
and (\ref{eq:I_slash}) to general relativity is problematic because the Einstein's 
equations are nonlinear in the background metric $g_{\mu\nu}$~\cite{MTW}. 
However, since in the present physical context no black hole QNMs excitation is
found, the nonlinear structure of the background metric, represented by the potential
$V^{\rm(e)}$, can be neglected in practice and the use of
the quadrupole formula can be justified a posteriori. Following a pragmatic 
approach, a reasonable modification to account for general relativistic effects 
can be obtained by projecting $t^{00}$ onto the local inertial frame of the tetrad 
associated with the stationary observers of the Schwarzschild spacetime, so that~$t^{\hat{0}
\hat{0}}=\alpha^2t^{00}$, where $t^{00}=\rho h u^0 u^0+pg^{00}$ and $\alpha=
\sqrt{-g_{00}}$ is the lapse. Here $p$ is the pressure, $h=1+\varepsilon+p/\rho$ 
is the specific enthalpy, and $\varepsilon$ is the specific internal energy of the fluid. 
Using spherical coordinates we obtain
\begin{align}
\label{eq:quad_polar}
I=2\pi\int_{-1}^{1}dz\int_{0}^{\infty} \left(hDW-p\right) \left(\frac{3}{2}z^2-\frac{1}{2}\right)
r^4 dr\;,
\end{align}
where $I\equiv 3\bI_{\hat{3}\hat{3}}/2 $, $z=\cos\theta$, $D=\rho W$ is the relativistic 
rest-mass density, and $W=\alpha u^0=(1-g_{ij}v^iv^j)^{-1/2}$ is the Lorentz 
factor. If we define $A^{E2}_{20}\equiv\sqrt{64\pi/15}\;d^2I/dt^2$, the 
energy emitted in the quadrupole approximation is~\cite{dimmelmeier}
\begin{equation}
\label{energy:SQF}
E=\frac{1}{32\pi^2}\int_0^{\infty}\omega^2|\tilde{A}^{E2}_{20}(\omega,r)|^2 d\omega\;.
\end{equation}
In the following we use SQF$_1$ to denote the standard quadrupole formula with 
the generalized quadrupole moment proposed above. Figure~\ref{label:fig2} 
displays the comparison, for model B, of the gravitational waveforms obtained 
with the Zerilli-Moncrief equation and with the SQF$_1$ definition for $\bI_{\hat{3}\hat{3}}$. 
The second time derivative has been computed directly from the numerical 
data using centered finite-differences. We note that for our particular
physical setup and unlike the case of core collapse simulations~\cite{finn90,dimmelmeier}, 
the numerical noise associated with this procedure is negligible. 
Fig.~\ref{label:fig2} shows an excellent agreement between the two methods: the two lines 
perfectly overlap at early times and only small differences are reached after 
many orbital periods. Fig.~\ref{label:fig3} 
shows the corresponding comparison of the energy spectra for $t=100t_{\rm orb}$ 
and (in the inset) a close up of the waveforms for different alternative 
definitions of $t^{\hat{0}\hat{0}}$. The Newtonian SQF ($t^{\hat{0}\hat{0}}=\rho$) 
slightly underestimates the wave amplitude while SQF$_2$ ($t^{\hat{0}\hat{0}}=\rho W$) 
is closer to the gauge-invariant signal; SQF$_3$, corresponding to 
$t^{\hat{0}\hat{0}}=\rho u^0$ and used in Refs.~\cite{blanchet90,shibata03,dimm04},
gives the worst relative agreement in the present physical context (dash-dotted line in 
the inset of Fig.~\ref{label:fig3}).

%--------------------------------------- fig 3 --------------------------------------------
\begin{figure}[t]
\begin{center}
\includegraphics[width=75 mm, height=65 mm]{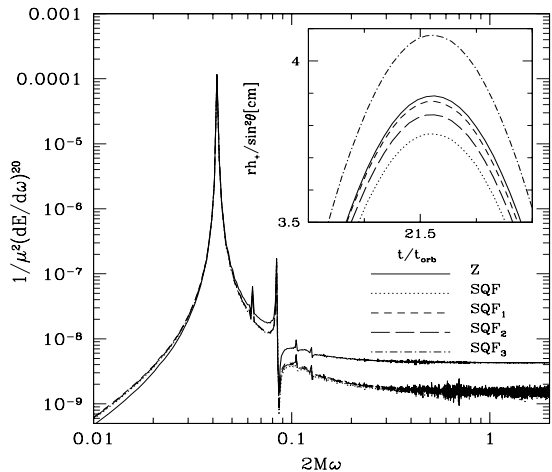}
\caption{Comparison of the energy spectra for model B and a portion of the waveform when different 
definitions of the generalized quadrupole moment of the torus are used.}
\label{label:fig3}
\end{center}
\end{figure}
%--------------------------------------- fig 3 --------------------------------------------

To quantify the differences among the various alternative expressions we use for the
quadrupole moment we report in Table~\ref{label:table2} the relative errors $\Delta E^{20}$ 
found in the total energy emitted after the frequency integration of the power spectra
for all four models of our sample. We recall that from model A to D the radial 
location of the center of the disk increases gradually. It is found that the SQF$_1$ 
(fourth column in Table~\ref{label:table2}) gives the best agreement with the 
gauge-invariant GW extraction, with an error that is always less than $2\%$. On 
the other hand, the SQF$_3$ yields the worst result for either model, systematically 
overestimating the amplitude of the waveforms, the reason being that, when compared
to the other expressions, the functional form of SQF$_3$ contains a division by an 
extra lapse function ($\alpha<1$). 

It needs to be mentioned, however, that the decrease in $\Delta E^{20}$ with increasing 
$r_{\rm center}$, observed in SQF and SQF$_3$ in Table~\ref{label:table2}, seems 
accidental. This has been checked by analyzing additional models with the 
same linear dimension but with the center located at $r_{\rm center}=10.17$
and $r_{\rm center}=10.47$. The latter is the maximum allowed value for a
closed disk with $l=4$~\cite{font02a}. No clear continuation in the trend in 
$\Delta E^{20}$ reported in Table~\ref{label:table2} has
been found for these additional models. The largest errors computed appear in
SQF$_3$ for the model with $r_{\rm center}=10.17$ ($\Delta E^{20}\sim 5.4\%$)
and in SQF for the model with $r_{\rm center}=10.47$ ($\Delta E^{20}\sim 5\%$).

Further evidence on the sensitivity of these results on the model properties is
gained by reducing the potential gap in model D to $\Delta W=-0.008$, keeping the 
center at the same location. As a result the disk becomes much larger, extending 
from $r_{\rm in}=4.93$ to $r_{\rm out}= 42.5$. In this case, the relative difference 
in the emitted energy between the gauge--invariant waveform and the quadrupole 
formula can be as large as $\sim 25\%$ for SQF, SQF$_1$ and SQF$_2$, while 
$\Delta E^{20}\sim 21\%$ for SQF$_3$. Following Tanaka {\it et al.}~\cite{tanaka93}, 
who computed GWs from particles orbiting around a Schwarzschild black hole using 
linear perturbation theory, we interpret this disagreement for enlarged tori as 
the result of GWs back-scattering with the tail of the background potential.

%----------------------------------- table 2 ---------------------------------------------
\begin{table}[t]
\caption{\label{label:table2}
Relative differences $\Delta E^{20}=|E^{20}-E|/E^{20}$ among the emitted energies computed using
different versions of the quadrupole formula ($E$) and the gauge-invariant approach ($E^{20}$)
for disk models with different position of the center.}
\begin{ruledtabular}
\medskip
\begin{tabular}{ccccccc}
Model        & $r_{\rm center}$ & $\Delta E^{20}_{\rm SQF}$  & $\Delta E^{20}_{{\rm SQF}_{1}}$ 
& $\Delta E^{20}_{{\rm SQF}_{2}}$ &$\Delta E^{20}_{{\rm SQF}_{3}}$  &\\
\hline
A  &  7.720 & 2.9$\%$  &  $0.7\%$  &  $0.8\%$  & $11.0\%$   &\\
B  &  7.720 & 5.9$\%$  &  $0.3\%$  &  $2.6\%$  & $10.6\%$ &  \\
C  &  8.352 & 4.9$\%$  &  $1.8\%$  &  $3.2\%$  & $6.7\%$  &   \\
D  &  9.971 & 2.7$\%$  &  $1.7\%$  &  $2.2\%$  & $3.8\%$  &                                  
\end{tabular}
\end{ruledtabular}
\end{table}
%------------------------------------------------------------------------------------------

As mentioned above, in simulations of core collapse~\cite{finn90,dimmelmeier}, 
the accurate numerical computation of the time derivatives appearing in the SQF is 
prone to introduce high-frequency numerical noise. The common procedure to
circumvent this is to rewrite Eq.~(\ref{eq:h_plus}) making use of the
Newtonian limit of the general relativistic hydrodynamics equations (i.e.~the Euler 
and continuity equations)~\cite{finn89,blanchet90} to eliminate the time derivatives. 
In axisymmetry the resulting expression reads
\begin{eqnarray}
\label{eq:stress_formula}
A^{E2}_{20}&=&k\int_{-1}^{1}dz\int_{0}^{\infty}\rho\bigg[v_r v^r (3 z^2 \!-\! 1)\nonumber\\
           &+& v_\theta v^\theta (2 \!-\! 3 z^2) \!-\!
        v_\varphi v^\varphi 
         -6 z \sqrt{(v^r v_r)(v_\theta v^\theta)
        (1 \!-\! z^2)} \biggl. \nonumber\\
        &-& r \partial_r\Phi
        (3 z^2 \!-\! 1)
        +\, 3 z\de_{\theta}\Phi
        \sqrt{1 \!-\! z^2}\bigg]r^2 dr \;,
\end{eqnarray}
where $k=16\pi^{3/2}/\sqrt{15}$ and $\Phi$ is the Newtonian gravitational potential 
generated by the central source $M$. This equation is known as the {\em stress formula} 
(SF), and it has been applied in diverse astrophysical 
scenarios~\cite{zanotti03,dimmelmeier,finn89}. Since Eq.~(\ref{eq:stress_formula}) is 
obtained from the Newtonian fluid equations, its application in a general relativistic 
code poses some ambiguities. However, it is possible to introduce suitable forms for 
$\Phi$ in Eq.~(\ref{eq:stress_formula}) to obtain results which are consistent with 
those derived in the perturbative framework.  
%
%------------------------------ FIGURE 4 ------------------------------------ 
\begin{figure}[t]
\begin{center}
\includegraphics[width=75 mm, height=65 mm]{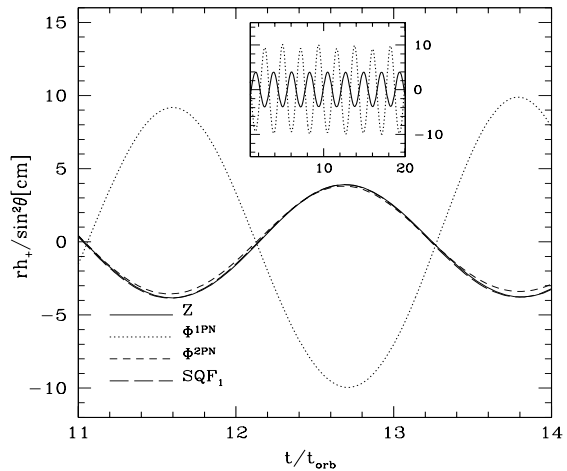}
\caption{Model B: comparison between the Zerilli-Moncrief function, the SQF$_1$ and 
the SF for different choices of the potential $\Phi$. The inset shows more periods 
of oscillations.} 
\label{label:fig4}
\end{center}
\end{figure}
%---------------------------------------------------------------------------
%
An elementary possibility is to use a first post-Newtonian (1PN) expression 
for $\Phi$, as it was done in~\cite{zanotti03}. In practice $\Phi$ can be 
obtained from $g_{rr}=1-2\Phi$ which is correct at 1PN~\cite{blanchet_llr}; 
i.e.~$\Phi\equiv\Phi^{\rm 1PN}=-M/(r-2M)$. Moreover, one can also introduce 
the first terms of the 2PN expansion of $g_{rr}$~\cite{blanchet_llr}, solving 
the quadratic equation $g_{rr}=1-2\Phi+2\Phi^2$ and choosing the solution 
which recovers the Newtonian potential $\Phi^{\rm N}=-M/r$ for large $r$; i.e, 
$\Phi\equiv\Phi^{\rm 2PN}=1/2\left[1-\sqrt{(r+2M)/(r-2M)}\right]$. 
Figure~\ref{label:fig4} shows the GW signals for model B computed using the 
SQF$_1$ (long-dashed line), the Zerilli-Moncrief function $Z$ (solid line), 
and the SF with $\Phi^{\rm 1PN}$ (dotted line) or $\Phi^{\rm 2PN}$ (short-dashed 
line). The waveforms obtained from the SF present a systematic global vertical 
offset~\cite{zanotti03} which was corrected in the figure. Figure~\ref{label:fig4} 
shows that a good agreement is found among all waveforms except for those 
computed using $\Phi^{\rm 1PN}$. In this case, the signal is out of phase by 
half a period, due to the $\de_r\Phi$ term being dominant in 
Eq.~(\ref{eq:stress_formula}) and thus responsible for a change of sign, while 
the amplitude is about a factor of two larger than that of the other waveforms. 
On the other hand, it is remarkable the good agreement in amplitude found among 
the SF waveform using $\Phi^{\rm 2PN}$ with both the Zerilli-Moncrief equation 
and the SQF$_1$. We finally note that the use of a potential consistent with 
the Schwarzschild metric, obtained from $g_{00}=\exp(-2\Phi)$, results in a 
signal whose amplitude is $\sim 30\%$ larger than the one obtained with SQF$_1$. 
This stresses the intrinsic ambiguity which relies in the use of a modified 
Newtonian formula in a general relativistic context. Work is in progress to 
derive a modified stress formula which holds in this regime.

In this paper we have studied the GW signals driven by oscillations of thick 
accretion disks around a Schwarzschild black hole by means of gauge-invariant 
metric perturbation theory and modifications of the SQF. The comparison between 
the two approaches used has been discussed in detail. For the models considered, 
the GW signal has been found to be mainly determined by the matter flows and 
no evidence of black hole QNM excitation has been observed.

We thank P.~Cerd\'a-Dur\'an, V.~Ferrari, L.~Gualtieri, J.A.~Pons, M.~Portilla, and 
L.~Rezzolla for useful discussions. This work has been supported by the Angelo Della Riccia 
Foundation, the Spanish MECD (SB2002-0128), and the Spanish MCyT (grant AYA 2001-3490-C02-01). 
The computations were performed on the Beowulf Cluster for Numerical Relativity 
{\tt Albert100}, at the University of Parma.
 
%%%%%%%%%%%%%%%%%%%%%%%%%%%%%%%%%%%%%%%%%%%%%%%%%%%%%%%%%%

\end{document}